\documentclass[12pt,a4paper]{article}
\usepackage{jheppub}
\usepackage[FIGTOPCAP]{subfigure}
\usepackage{amsmath}
\usepackage{amsfonts}
\usepackage{amssymb}
\usepackage{tensor,epigraph}
\usepackage{cancel}
\usepackage{ulem} 
\usepackage{musixtex}
\usepackage{booktabs}
\usepackage{empheq}
\usepackage{amsmath}
\usepackage{amssymb}
\usepackage{amsthm}
\usepackage{comment}
\usepackage{psfrag}
\usepackage{graphicx}
\usepackage{color}
\usepackage{bbm}
\usepackage{color}

\newcommand{\be}{\begin{equation}}
\newcommand{\ee}{\end{equation}}

\definecolor{klgreen}{rgb}{0.0, 0.5, 0.0}

\newcommand{\exclude}[1]{}

\newcommand{\beq}{\begin{equation}}
\newcommand{\eeq}{\end{equation}}
\newcommand{\bea}{\begin{eqnarray}}
\newcommand{\eea}{\end{eqnarray}}
\long\def\/*#1*/{}

\newcommand{\junk}[1]{}

\title{Multi-field Cuscuton Cosmology}
\author{Seyed Ali Hosseini Mansoori $^{a}$\footnote{shosseini@shahroodut.ac.ir},}
\author{ and Zahra Molaee $^{b}$\footnote{zmolaee@ipm.ir}, ~~~~~~~~~~~~~~}
\affiliation[a]{Faculty of Physics, Shahrood University of Technology, P.O. Box 3619995161 Shahrood, Iran}
\affiliation[b]{School of Astronomy, Institute for Research in Fundamental Sciences (IPM), P.O. Box 19395-5531, Tehran, Iran\vspace{0.1cm}}

\vspace{0.1cm}

\abstract{ In this paper, we first introduce a multi-field setup of Cuscuton gravity in a curved field space manifold. Then, we show that this model allows for a regular bouncing cosmology and it does not lead to ghosts or other instabilities at the level of perturbations. More precisely,
 by decomposing the scalar fields perturbations into the tangential and normal components with respect to the background field space trajectory, the entropy mode perpendicular to the background trajectory is healthy which directly depends on the signature of the field-space metric, whereas the adiabatic perturbation tangential to the background trajectory is frozen. In analogy with the standard Cuscuton theory equipped with an extra dynamical scalar field, the adiabatic field does not have its own dynamics, but it modifies the dynamics of other dynamical fields like entropy mode in our scenario. Finally, we perform a Hamiltonian analysis of our  model in order to count the degrees of freedom propagated by dynamical fields. }
\preprint{
}

\begin{document}
\maketitle
%\allowdisplaybreaks

%%%%%%%%%%%%%%%%%%%%%%%%%%%%%%%%%%%%%%%%%%%%%%%%%%%%%
\section{Introduction}\label{sec0}

In recent years, there has been growing interest in modified gravity theories which can be used to explain some of the unsolved problems in cosmology and General Relativity (GR) such as  unknown components of universe called the dark matter and dark energy, the singularity problem and so on.

%General relativity (GR) is successful in describing many physical phenomena.  However, it can not explain an accelerated expansion of the universe, dark matter and dark energy. Thus, 
%the unknown parts of the universe due to new physics beyond GR.  With regard to this point, some modification  is essential for GR in which during the past several years, many modified gravity models have been proposed  .

Most modified gravity theories such as Horndeski and Galileon models \cite{Horndeski:1974wa, Deffayet:2011gz, Zumalacarregui:2012us,Gao:2011qe, Gleyzes:2014dya, Gleyzes:2014qga,Nicolis:2008in} contain extra degrees of freedom governed by dynamical equations. In spite of this fact, recently the Cuscuton gravity \cite{Afshordi:2006ad, afshordi2007cuscuton} has been proposed as an infrared modification of GR, with no additional degree of freedom. Cuscuton gravity can be implemented by adding a non-canonical scalar field to general relativity. In this sense, the equation of motion of this field does not have any second order time derivatives. It means that the Cuscuton field does not have its own dynamics and acts as an auxiliary field. Therefore, in order to produce dynamics, it is required to include other fields \cite{Afshordi:2006ad}. An increasing number of studies have been carried out on Cuscuton gravity and its extended versions. As an example, in \cite{Afshordi:2009tt, bhattacharyya2018revisiting} the authors proved that   Cuscuton model with a quadratic potential can be considered as a low-energy limit of the non-projectable Horava-Lifshitz gravity model  \cite{hovrava2009quantum}. 
 Further, such a theory provides significant new features such as a distinct Cosmic Microwave Background \cite{afshordi2007cuscuton}, viable power-law solutions of inflation \cite{ito2019dressed}, the absence of spherical caustic instabilities in galileon theories \cite{de2017caustics}, and inflationary solutions \cite{adshead2016magnon, bartolo2022cuscuton}. Interestingly, in Ref. \cite{DeFelice:2022uxv} the authors have also shown that all acceptable Cuscuton solutions are always solutions for VCDM theory, which is a kind of Type-IIa Minimally Modified Gravity theories \cite{de2020theory}. 
 
 Remarkably, it was shown recently that a stable regular bouncing cosmology generated by Cuscuton gravity
 \cite{boruah2018cuscuton,quintin2020cuscuton,kim2021spectrum}. Generally, regular bouncing models violate the Null Energy Condition (NEC), which results in either instabilities or a superluminal speed of sound \cite{dubovsky2006null,rubakov2014null,sawicki2013hidden,libanov2016generalized,kobayashi2016generic}\footnote{The stability and superluminal nature of the bouncing models is still
a challenging problem in the literature. For an example, in \cite{Ijjas:2016vtq, Ijjas:2016tpn, Creminelli:2012my} authors have argued that a healthy bounce can be achieved using Galilean action,  whereas in \cite{Dobre:2017pnt, Easson:2013bda, Easson:2011zy} have been shown that when coupling to matter or other regions of phase-space are included, Galilean models have superluminal speed of sound.}.
 To be more explicit, due to the non-dynamical nature of the Cuscuton field,  an effective violation of the NEC occurs for the background bounce while the actual dynamical degree of freedom does not violate NEC and remains safe. Furthermore,  in \cite{boruah2017theory,boruah2018cuscuton} the authors have shown that a Cuscuton bounce does not suffer from ghost instabilities and the scalar perturbations remain stable throughout the bounce phase. The stability of these perturbations were investigated in further detail for (generalized) Cuscuton setups in Refs. \cite{quintin2020cuscuton,iyonaga2018extended}. 
 
   With all this in mind, it is interesting to ask what happens if one considers a multi-field generalization of Cuscuton gravity. In the last few years there has been a growing interest in multi-field models with curved field-space manifold, ranging from inflation, dark energy, primordial non- Gaussianity and related areas \cite{gong2011covariant,gong2017multi,gong2016path,cespedes2012importance,achucarro2017cumulative}. Therefore, in this work we first attempt to extend the single field Cuscuton gravity \cite{Afshordi:2006ad} to the multi-field setup with a curved field space manifold. Then, by projecting equations of motion for scalar fields along the tangent vector to the background field trajectory, we show that the adiabatic combination of scalar fields does not have its own dynamics, while it can modify the dynamics of the other dynamical fields in our scenario. In this respect, we can investigate the possibility of a bounce solution in such a model.

   At the level of pertubations, after decomposing  scalar fields perturbations into the tangential and normal components with respect to the background field space trajectory, finally we will prove that the multi-field Cuscuton bounce does not have any ghost instabilities and the scalar perturbations are stable throughout the bounce phase.  In addition,  inspired by Ref. \cite{gomes2017hamiltonian}, which the Hamiltonian analysis of the single field Cuscuton theory has been performed, we consider the full non- linear Hamiltonian analysis of the multi-field Cuscuton gravity with the curved field space manifold and calculate the physical degrees of freedom.

The paper is organized as follows. In Section \ref{bb},  we attempt to construct a generalization of the multi-field 
 Cuscuton model with non-canonical kinetic term. Moreover, we derive  the background equation of motions, then we obtain equation of motions along orthogonal and tangent unit vectors to the field trajectory. 
 In Section \ref{sect3},  we find a Cuscuton bounce solution for our model. 
Cosmological perturbations are presented in Section \ref{sec4}, and quadratic action in the spatially flat gauge and comoving gauge are considered. 
 Section \ref{sec5} is devoted to a Hamiltonian formalism of the multi-field Cuscuton gravity in non-linear level.     
 Finally, our conclusions are drawn in Section \ref{sec7}.

%%%%%%%%%%%%%%%%%%%%%%%%%%%%%%%%%%%%%%%%%%%%%%%%%%%%%
 \section{Building a model for the multi-field Cuscuton gravity} \label{bb}
Let us first consider a multi-field system with a generic field space metric $G_{ab}$ coupled to the Einstein gravity which is given by \cite{langlois2008perturbations} 
\begin{equation}\label{action1}
S=\int d^4 x \sqrt{-g} \Big(\frac{M_{p}^2}{2}R+P(X)-V(\phi^{a})\Big),
\end{equation}
in which $M_{p}$ is the reduced Planck mass, $R$ is the Ricci scalar associated with the spacetime metric $g_{\mu\nu}$ and $P$ is an arbitrary function of the kinetic term $X=- G_{ab} g^{\mu \nu} \partial_{\mu} \phi^{a}\partial_{\nu} \phi^{b}/2$ \footnote{Generally, one can take $P$ as a function of $X$ and $\phi^{a}$. We here assume $P$ depends only on $X$, for simplification. }.  The $V$ is a general potential function depending on the scalar fields $\phi^{a}$ as well.
By varying the action (\ref{action1}) with respect to the metric $g_{\mu \nu}$, the energy-momentum tensor is obtained to be
\begin{equation}\label{EMT1}
T_{\nu}^{\mu}=(P-V)\delta_{\nu}^{\mu}+P_{,X}G_{ab}\partial_{\nu}\phi^{a} \partial^{\mu} \phi^{b},
\end{equation}
in which $P_{,X}$ denotes the partial derivative of $P$ with respect to $X$. In a spatially flat FLRW spacetime,
\begin{equation}
ds^2=-dt^2+a(t)^2 \delta_{ij}dx^{i}dx^{j},
\end{equation}
the energy-momentum tensor (\ref{EMT1}) reduces to that of a perfect fluid with the energy density 
\begin{eqnarray}
\rho=2 X P_{,X}-p,
\end{eqnarray}
and pressure $p=P-V$. The background equation of motion for the scalar field can be obtained by the variation of the action (\ref{action1}) with respect to $\phi^{a}$ which yields \cite{langlois2008perturbations}
\begin{equation}\label{EOM3}
D_{t} \dot{\phi}^{a}+\Big(3 H +\frac{\dot{P}_{,X}}{P_{,X}}\Big) \dot{\phi}^{a}+\frac{1}{P_{,X}}G^{ab}V_{,b}=0,
\end{equation}
where $_{,b}$ stands for the derivative with respect to the scalar field $\phi^{a}$ and  $D_{t}$ is the covariant time derivative which is defined as
\begin{equation}
D_{t} A^{a}\equiv \dot{A}^{a}+\Gamma^{a}_{bc}A^{b} \dot{\phi}^{c},
\end{equation}
where $\Gamma_{bc}^{a}$ is the Christoffel symbol constructed by the field space metric $G_{ab}$ and $D_{t}G_{ab}=0$. 
Because $\phi^{a}$ are coordinates in the field space, we are allowed to choose other convenient basis. One possible choice is the so-called kinematic basis which is a set of orthogonal unit vectors $\{e_{n}^a\}$ ($n=1,..,\mathfrak{N}$) on the $\mathfrak{N}$- dimensional field space, where the first vector is the unit \textit{adiabatic} vector  defined as
\begin{eqnarray}&&\label{eq1NB}
e_{1}^{a} \equiv T^a \equiv \frac{\dot \phi_0^a}{\dot \phi_{0}}
\end{eqnarray}
in which $\dot{\phi_0^2}=G_{ab}\dot \phi_{0}^{a} \dot \phi_{0}^{b}=2 X$. The $(\mathfrak{N}-1)$ remaining vectors of the basis, $N_{n}^{a}$, span the entropy subspace which is orthogonal to the adiabatic direction \cite{gong2011covariant,Elliston:2012ab}. One can also define useful matrices as $Z_{mn}\equiv e_{ma}D_{t}e^{a}_{n}$ which satisfy the antisymmetry property $Z_{nm}=-Z_{mn}$ as a result of  
 $D_{t}(e^a_{m}e_{na})=0$ \cite{gong2011covariant,Elliston:2012ab}. %More precisely, the normal vectors $N_{n}^{a}$ not only satisfy the condition $G_{ab}N_{n}^{a}T^{a}=0$, but also are naturally proportional to the derivative of $T^{a}$, that is $N^{a} \propto D_{t} T^{a}$ \cite{gong2011covariant,Elliston:2012ab}. 

%These two unit vectors may be used to decompose the scalar field equation of motion  (\ref{EOM3}) into adiabatic and entropic equations.

Therefore, by projecting Eq. (\ref{EOM3}) along $T^a$, the adiabatic equation of motion is obtained to be
\begin{equation}\label{adiabaticeq1}
\ddot{\phi_{0}}+\Big(3 H +\frac{\dot{P_{,X}}}{P_{,X}}\Big)\dot{\phi}_{0}+\frac{1}{P_{,X}}V_{T}=0,
\end{equation}
where $V_{T}\equiv  V_{,a} T^{a}$. In addition, by using 
\begin{equation}
\dot{P}_{,X}=P_{,XX}\dot{X}=P_{,XX} \dot{\phi}_{0} \ddot{\phi}_{0},
\end{equation}
Eq. (\ref{adiabaticeq1}) can be written as \cite{langlois2008perturbations} 
\begin{equation}\label{adibaticEOM}
(P_{,X}+2X P_{,XX})\ddot{\phi_{0}}+3 H P_{,X} \dot{\phi}_{0}+V_{T}=0.
\end{equation}
On the other hand, the entropy part of the equation of motion gives us the rate of change of the adiabatic (tangent) vector $T_{a}$. Thus,
by projecting Eq. (\ref{EOM3}) along $N_{n}^a$, we also arrives at 
\begin{equation}
D_{t}T^a=-\frac{V_n}{P_{,X} \dot{\phi}_{0}}N_{n}^a, \label{time-deriv-T}
\end{equation}
where $V_n = N_{n}^a V_{,a}$. This also implies that the matrix elements $Z_{n1}=-Z_{1n}=-V_n/(P_{,X} \dot{\phi}_{0})$.
 In analogy with the single field cuscuton gravity \cite{Afshordi:2006ad,afshordi2007cuscuton,boruah2017theory,boruah2018cuscuton,kim2021spectrum},  we therefore demand that a multi-field generalization of Cuscuton gravity can be achieved by taking  the below constraint
\begin{equation}\label{cons1}
P_{,X}+2X P_{,XX}=0.
\end{equation}
Obviously, in this limit, the adiabatic equation of motion \eqref{adibaticEOM} is not second order due to the absence of the time derivative of $\dot{\phi}_{0}$. It means that the Cuscuton adiabatic field does not have its own dynamics, while it may modify the dynamics of the other dynamical fields such as entropy mode in our scenario (we refer readers to Section. \ref{sec4} in more detail.). 
%Note that in the original single field cuscuton gravity,  the cuscuton field acts as an auxiliary field without its own dynamical degree of freedom, whereas each scalar field is a dynamical field in our setup. Therefore, in order to induce a dynamical cosmological background one needs to include other scalar fields.
  Moreover, by imposing the above constraint on the speed of sound, 
\begin{equation}
c_{s}^2\equiv \frac{p_{,X}}{\rho_{,X}}=\frac{P_{,X}}{P_{,X}+2X P_{,XX}},
\end{equation}
this generality leads to superluminal speed of sound which is naturally addressing the violation of Null Energy Condition (NEC) in regular bounce models (we refer to this point in the next section.). % To make our setup, we add the constraint (\ref{cons1}) to the action (\ref{action1}) through a Lagrange multiplier $\lambda$ in the following form.
%\begin{equation}\label{cusaction}
%S=\int d^4 x \sqrt{-g} \Big(\frac{M_{p}^2}{2}R+P(X)-V(\phi^{a})+\lambda (P_{,X}+2X P_{,XX})\Big)
%\end{equation}
Note that a superluminal propagation speed may not necessary indicates that causality is violated \cite{babichev2008k} (we discuss about this fact in appendix \ref{appendixa}).

\section{Background Cosmology and bounce solutions}\label{sect3}
In this section, we consider the background equations of motion for multi-field Cuscuton model and present a
model for a multi-field Cuscuton bounce scenario. 
Similar to the single-field cuscuton gravity, the constraint \eqref{cons1} leads to restrict our choice of $P$ to
\begin{equation}\label{Pfunction}
P(X)=\pm \mu^2 \sqrt{2X},
\end{equation}
where $\mu$ is constant. It follows that Eq. (\ref{EOM3}) reduces to
\begin{equation}
\mp \mu^2  G_{ab} D_{t}\Big(\frac{a^3}{\sqrt{2X}}\dot{\phi}^{b} \Big)=a^3 V_{,a}.
\end{equation} 
It seems that scalar fields are dynamical fields, in contrast with the single field Cuscuton model in which the cuscuton field has not its own dynamical degree of freedom \cite{Afshordi:2006ad,afshordi2007cuscuton,boruah2017theory,boruah2018cuscuton,kim2021spectrum}. 
Nevertheless, the adiabatic combination of scalar fields is a field with no dynamics. It means that the adiabatic equation of motion (\ref{adiabaticeq1}) converts to
\begin{equation}\label{Eq33}
\pm 3 \mu^2  \text{sign}(\dot \phi_{0}) H+V_{T}=0,
\end{equation}
without second time derivative of the adiabatic field $\phi_{0}$. Moreover, Friedman equations are obtained to be
\begin{eqnarray}
3 M_{p}^2 H^2&=&V ,\label{FRI1}\\
M_{p}^2\dot{H}&=&-X P_{,X}=-\frac{1}{2} (\pm) \mu^2 \sqrt{2 X}\label{FRI2}.
\end{eqnarray} 

Now allow us to investigate the bounce realization in the framework of multi-field Cuscuton cosmology.
An important feature of a regular bounce ($H\neq  \pm \infty $) is that universe moves from a contracting phase ($H<0$) into an expanding phase ($H>0$) at finite value of the scale factor $a_{b}$. It follows that
\begin{equation}
H_{b}=0, \hspace{0.5cm} \text{and} \hspace{0.5cm} \dot{H}_{b}>0.
\end{equation}
Obviously, the second condition implies the violation of NEC. Therefore, this condition forces us to consider the negative sign for the adiabatic field in Eq. \eqref{FRI2} to get a bounce solution. This in turns, yields Eq. \eqref{Eq33} becomes

\begin{equation}\label{EqqVT}
- 3\mu^2  \text{sign}(\dot \phi_{0}) H+V_{T}=0.
\end{equation}
Without loss of generality, from now on, we only consider solutions with $\dot{\phi}_{0}>0$. Taking a time derivative from both sides of Eq. (\ref{EqqVT}), we have
\begin{equation}
3 \mu^2 \dot{H}=V_{TT} \dot{\phi}_{0}+\Big(\frac{V_{n}}{\mu}\Big)^2.
\end{equation}
where $V_{n}=N_{n}^{a}V_{,a}$ and $V_{TT}=T^{a}T^{b}V_{;ab}$. Because the last term in the right hand side of the above equation is always positive, one can consider
\begin{equation}
V_{TT}>0.
\end{equation}
near to the bounce where the NEC will be violated ($\dot{H}>0$).
Moreover, by plugging $H$ from Eq. (\ref{EqqVT}) back into Eq. (\ref{FRI1}), one obtains
\begin{equation}\label{EV2}
\frac{M_{p}^2}{3 \mu^{4}} V_{T}^2=V,
\end{equation}  
taking a time derivative of the above equation, we arrive at
\begin{equation}\label{EV1}
\frac{2M_{p}^2}{3 \mu^4}V_{TT}-1=-\frac{2M_{p}^2}{3 \mu^4 \dot{\phi_{0}}}\Big(\frac{V_{n}}{\mu}\Big)^2<0,
\end{equation}
thus
\begin{equation}
V_{TT}<\frac{3 \mu^{4}}{2 M_{p}^2}.
\end{equation}
As a result, while the shape of the potential in NEC violation along tangent basis is convex ($V_{TT}>0$), its convexity is in the below range
\begin{equation}
0<V_{TT}<\frac{3 \mu^4}{2 M_{p}^2}.
\end{equation}
This allows that universe experiences a regular bouncing cosmologies, where an initially contracting universe, bounces and starts expanding.
With the background quantities established, we can study cosmological perturbations and analyze the existence of ghosts and other instabilities in this model.

\section{Cosmological Perturbations }\label{sec4}

%The next step is to survey cosmological perturbations  with help of the background quantities in our model.

 In this part, we investigate the dynamical stability of the cosmological scalar perturbations in multi-field Cuscuton gravity. The set up we used here is inspired by the one proposed in Ref. \cite{gong2011covariant,Elliston:2012ab}.  
 %Scalar fields $\phi^{a}(x^{\mu})$  break down into the homogeneous background value, $\phi^{a}_{0}$, and the gauge dependent fluctuations, $\delta \phi^{a}$ in space time.  The fluctuations $\delta \phi^{a}$ show a finite coordinate displacement from the classical trajectory and they are not covariant. This due to  the vector fields  $Q^{a}$ in order to set down the field fluctuations in a covariant form.  
%The two points like $\phi_0^a(t)$ and $\phi^a = \phi_0^a + \delta\phi^a$ are associated with a unique geodesic as regards the field space metric $G_{ab}$  \cite{gong2011covariant,Elliston:2012ab}
%where this geodesic is parameterized by $\varepsilon$, such that $\phi^{a}(\varepsilon=0)=\phi^{a}_{0}$ and $\phi^{a}(\varepsilon=1)=\phi^{a}_{0}+\delta \phi$. 
%These boundary conditions obtain a unique vector $Q^{a}$ which connects the two scalar field values in such a way that $\left. D_\varepsilon\phi^a \right|_{\varepsilon=0} = Q^a \ $ where $D$ is 
%the covariant derivative with respect to the field space metric $G_{ab}$. 
At the perturbation level, we assume $\phi^a = \phi_0^a + \delta\phi^a$ where $\delta \phi^a$ are the scalar perturbations. Similar to the Riemann normal coordinates \cite{trofimov2002riemannian,petrov1969phys,muller1997closed}, these perturbations can be expressed as a power series of the tangent vector $Q^{a}$ to the geodesic trajectory in the field space. Therefore, one can cosider the below covariant form for scalar perturbations  \cite{gong2011covariant,Elliston:2012ab}. 
\begin{equation}
\label{eq:mapping2}
\delta\phi^a = Q^a - \frac{1}{2}\Gamma^a_{bc}Q^bQ^c + \frac{1}{6} \left( \Gamma^a_{de}\Gamma^e_{bc} - \Gamma^a_{bc;d} \right) Q^bQ^cQ^d + \cdots \, .
\end{equation}
in which $\Gamma^a_{bc}$ represents the Christoffel symbol associated with the metric $G_{ab}$. Clearly,  the field fluctuations $\delta\phi^a$ and $Q^a$ are identical at the linear order, but they are different at higher orders. Furthermore, components of the full metric denote via $g_{00}= -{\mathcal N}^2 + \beta_i \beta^i, g_{0i}= \beta_i, g_{ij} = \gamma_{ij}$ in which at linear order the scalar perturbation parts are given  by
\begin{align}\label{pertubation}
\nonumber \mathcal N & = 1 + \alpha \, ,
\\
\beta_i & =B_{,i}  \, ,
\\
\nonumber \gamma_{ij}& =a^{2} e^{2 \psi} \delta_{ij}
\end{align}
where $\mathcal{N}$ is the lapse function while $\beta_{i}$ is the shift vector. In spatially flat gauge, we take $\psi=0$
and thus $\gamma_{ij} = a^2 \delta_{ij}$.

\subsection{Quadratic action in the spatially flat gauge
}

Let us first verify the cosmological perturbations in the spatially flat gauge. 
According to our results in previous section, the action \eqref{action1} is given by the following form\footnote{In Ref. \cite{adshead2016magnon}, the authors proposed a class of multi-scalar effective field theories (EFTs) that can achieve inflationary solutions. Remarkeably, this EFT superficially resembles the muli-field Cuscuton model at low energies.}   
\begin{equation}\label{22}
S=\int  d^4 x  \sqrt{-g}\big(\frac{M^2_{p}}{2}R-\mu^{2}\sqrt{2X} -V(\phi^{a})\big),
\end{equation}
%Before proceeding further, let us examine the impact of the constraint contribution term in (\ref{cusaction}) on the quadratic action. At the second order of perturbations, one obtains   
%\begin{eqnarray}
%\nonumber \delta^2\Big(\sqrt{-g} \lambda f(X)\Big)&=&\sqrt{-g}  f_{X} \delta \lambda \delta X+\sqrt{-g} \lambda f_{,XX} \delta^2 X+\sqrt{-g} f \delta^2 \lambda+\lambda f_{X} \delta X \delta(\sqrt{-g})\\
%&+&f \delta \lambda  \delta(\sqrt{-g})+\lambda f\delta^2(\sqrt{-g})
%\end{eqnarray}  
%Where $f=P_{,X}+2XP_{,XX}$. Because the constraint (\ref{cons1}) is held for each order of derivative with respect to $X$, i.e. $f=0$, $f_{X}=0$, and $f_{XX}=0$, the above contribution will be vanished in the quadratic action. 
Now by expanding the above action up to the second order of scalar perturbations,
the quadratic action takes the following form,
\begin{eqnarray}\label{actionQ1}
&&S^{(2)}=\int d^4x \frac{a^{3}}{2}\Big[P_{,X} D_{t}Q_{b}D_{t}Q^{b}-\frac{1}{2X}P_{,X}D_{t}Q_{b}D_{t}Q_{c}\dot{\phi}^{b}\dot{\phi}^{c}- a^{-2} G_{bc}P_{,X}\partial_{i}Q^{c}\partial^{i}Q^{b}\\\nonumber  &&-2\alpha\big(3H^2 \alpha+ Q^{b}V_{b}+2 M_{p}^2H\partial_{i}\partial^{i}B \big)+2Q^{b}\dot{\phi}_{b}P_{,X}\partial_{i}\partial^{i}B -
Q^{b}Q^{c}\big(V_{bc}+ \mathbb{R}_{bdcf}\dot{\phi}^{d}\dot{\phi}^{f}P_{,X} \big)\Big],
\end{eqnarray}
where $V_{ab}=V_{;ab}$ and 
\begin{equation}
\mathbb{R}^{a}_{bd,c}\equiv \Gamma^{a}_{bd,c}-\Gamma^{a}_{bc,d} +\Gamma^{a}_{ce}\Gamma^{e}_{bd}-\Gamma^{a}_{de}\Gamma^{e}_{bc} \, ,
\end{equation}
is the Riemann tensor related to  the curved field space manifold.
Clearly, one can see that the quadratic Lagrangian is linear in terms
of the non-dynamical mode $ B $,  thus its equation of motion yields
\begin{equation}
\alpha=\frac{Q^{b}\dot{\phi}_{b}P_{,X} }{2 M_{p}^2 H} .
\end{equation}
Inserting the $ \alpha $ relation  in action \eqref{actionQ1}, the reduced action takes the following form
\begin{equation}\label{actionQ}
S^{(2)}=\int d^4x \frac{a^3}{2}\Big(P_{,X} D_{t}Q_b D_{t} Q^b - M_{bc} Q^{b}Q^{c}  -\frac{P_{,X}D_{t}Q_{b}D_{t}Q_{c}\dot{\phi}^{b}\dot{\phi}^{c}}{2 X}- \frac{G_{bc}}{a^2}P_{,X} \partial_{i}Q^{c}\partial^{i}Q^{b}\Big), 
\end{equation}
where the effective mass matrix is  introduced by
\begin{equation}\label{massterm}
M_{bc}=V_{bc}+\frac{P_{,X}}{2H}(V_{c}\dot{\phi}_{b}+V_{b}\dot{\phi}_{c})+\frac{3}{2 M_{p}^2}P_{,X}^{2}\dot{\phi}_{b}\dot{\phi}_{c}+P_{,X}\mathbb{R}_{bdcf}\dot{\phi}^{d}\dot{\phi}^{f}.
\end{equation} 
Up to now, our analysis was general, valid for any number of fields. 
We now consider explicitly the case where only two scalar fields are present. In this context, the entropy subspace is one-dimensional and the basis $\{e_{1}^{a},e_{2}^{a}\}=\{T^{a},N^{a}\}$ is completely determined. In this case, one can choose to write the normal one-form as
\begin{eqnarray}&&\label{eq1}
N_a = \left( sgn(\pm 1) G \right)^{1/2} \epsilon_{a b} T^{b} ,
\end{eqnarray}
in which $G$ is determinant of the metric $G_{ab}$, the signum function $sgn(\pm 1)$ determines the signature of $G_{ab}$, for instance, $sgn(-1)$ is for Lorentzian signature, whereas $sgn(+1)$ is chosen for Euclidean signature. In addition, $\epsilon_{a b}$ is the two dimensional Levi-Civita symbol with $\epsilon_{11} = \epsilon_{22} = 0$ and $\epsilon_{12} = - \epsilon_{21} = 1$. These definitions satisfy that $T_a T^a = 1$, $ N_a N^a=sgn(\pm 1)$ and $T^a N_a = 0$ \cite{gong2011covariant,Elliston:2012ab}. \\
By making use of Eq. (\ref{Pfunction}) and  (\ref{time-deriv-T}), we also conclude the following relations,
\begin{eqnarray}
D_{t}T^a& =&  \frac{V_{N}}{\mu^2} N^a=\dot{\theta} N^a , \\
D_{t}N^a& =& -\dot{\theta} T^a .
\end{eqnarray}
when $\dot{\theta} = 0$, the vectors $T^a$ and $N^a$ remain covariantly constant with respect to $D_t$ along the trajectory in the field space. If  $\dot{\theta}> 0$, the path turns to the left, whereas if $\dot{\theta} < 0$, the turn is towards the right. 
Moreover, the parallel and normal perturbations due to the background trajectory are given respectively by
\begin{align}\label{TDEF}
u^T & \equiv Q^T \equiv T_aQ^a \, ,
\\
u^N & \equiv Q^N \equiv N_aQ^a \, .
\end{align}
In this sense, $u^{T}$ corresponds to the perturbations parallel to the background trajectory which shows the adiabatic perturbation mode and $u^{N}$ equals to perturbations normal to the trajectory which indicates the entropy perturbation mode \cite{Gordon:2000hv, gong2011covariant,Elliston:2012ab}.  
By replacing $\dot \phi_{0}^{c}$ and $D_{t} T^{c}$ by the tangent and normal vectors $T^{c}$ and $N^{c}$ according to Eqs. (\ref{eq1}) and (\ref{TDEF}), and using the above representations, the quadratic action (\ref{actionQ}), finally reads
\begin{eqnarray}\label{Qadraticspa}
\nonumber S&=&\int d^4x \frac{a^3}{2} \Big[sgn(\pm 1)P_{,X} \Big(  (\dot{u}_{N})^2 -\frac{1}{a^2} (\partial u_{N})^2\Big)+sgn(\pm 1) P_{,X}\Big( \dot{\theta}^2 u_{T}^{2}-\frac{1}{a^2} (\partial u_{T})^2\Big)  \\ &+&2 sgn(\pm 1) P_{,X} \dot{\theta} u_{T}\dot{u}_{N}- M_{NN} u_{N}^{2}-2 M_{NT} u_{N}u_{T}-M_{TT}u_{T}^{2}\Big],
\end{eqnarray}
where
the symmetric matrix $M_{IJ}$ elements are specified by
\begin{eqnarray}
M_{NN}^2&=& N^{a}N^{b}M_{ab}^2= V_{NN}+sgn(\pm 1) \dot{H} \mathbb{R}\,,
\\
M_{TT}^2&=& T^{a}T^{b}M_{ab}^2=V_{TT}-\frac{\mu^2}{M_{p}^2 H}V_{T} +\frac{3}{2}\Big(\frac{\mu^4}{M_{p}^2}\Big)
\,,
\\
M_{NT}^2&=&M_{TN}^2= T^{a}N^{b}M_{ab}^2=V_{NT}-\frac{\dot{\theta} \mu^4}{2M_{p}^2 H}  \,,
\end{eqnarray} 
where $V_{NT}=N^{a}T^{b}V_{;ab}$, $V_{NN}=N^{a}N^{b}V_{;ab}$, and $V_{TT}=T^{a}T^{b}V_{;ab}$. Since we assume a 2D field space here, the Riemann tensor can be written in the terms of the Ricci scalar $\mathbb{R}$ as
\begin{equation}
\mathbb{R}_{abcd}=\frac{1}{2} \mathbb{R} \Big(G_{ac}G_{bd}-G_{ad}G_{cb}\Big),
\end{equation}
%hence, note the effect of the Ricci scalar $\mathbb{R}$ associated to the  field space  manifold. 

According to the quadratic action (\ref{Qadraticspa}), it is obvious that  $u_{N}$, \textit{i.e}.  the perturbation mode perpendicular to background trajectory is excited while the perturbation mode tangential to background trajectory, $u_{T}$, does not propagate. Particularly, the entropy mode propagates with a non-zero sound speed, although the sound speed for the adiabatic mode is zero.
Furthermore, whether the entropy perturbation is free from the gradient as well as ghost instabilities  depends on the signature of the metric, \textit{i.e}. $sgn(\pm 1)$. The similar result has been reported for the multi-field Mimetic gravity \cite{Mansoori:2021fjd}. 
Since at the bounce scenario 
\begin{equation}
\dot{H}=-\frac{X P_{,X}}{M_{p}^2}>0 \Rightarrow X P_{,X}<0\Rightarrow P_{,X}<0,
\end{equation}
in the case of the field space with an Lorntzian signature ($sgn(-1)=-1$), the entropy mode is healthy, whereas in the case of the Euclidean manifold with $sgn(+1)=1$, the entropy perturbation is pathological.  Therefore, we find that the Cuscuton bounce does not suffer from any ghost instabilities and the scalar perturbations remain stable throughout the bounce phase. 
In the next section, we confirm this finding in the comoving gauge as well.

\subsection{Quadratic action in the comoving gauge}\label{AppB}

In comoving gauge for the scalar perturbations,  $\psi$ is present  in \eqref{pertubation}. Thus, the equation of motion for the non-dynamical mode $B$ leads to the following constraint,
\begin{equation}
\alpha=\frac{\dot{\psi}}{H}+\frac{Q^{b}\dot{\phi}_{b}P_{,X} }{2 M_{p}^2 H} ,
\end{equation}
if we impose the above relation in the corresponding quadratic action, we will have the below quadratic action of the form
\begin{eqnarray}\label{comovingac}
\nonumber  S^{(2)}&=&\int d^4x \frac{a^3}{2}\Big(P_{,X} D_{t}Q_b D_{t} Q^b - M_{bc} Q^{b}Q^{c}  -\frac{P_{,X}D_{t}Q_{b}D_{t}Q_{c}\dot{\phi}^{b}\dot{\phi}^{c}}{2 X}- \frac{G_{bc}}{a^2}P_{,X} \partial_{i}Q^{c}\partial^{i}Q^{b}\\
&-&\dot{\psi} \frac{Q^{b} V_{b}}{2 H}+\frac{3}{2}\psi\Big[D_{t}Q_{b}\dot{\phi}^{b} P_{,X}-Q^{b}V_{b}\Big]-\frac{P_{X}}{2 a^2 H}\Big[Q_{b}\dot{\phi}^{b}\partial^2 \psi+\frac{X}{H}(\partial \psi)^2\Big]\Big),
\end{eqnarray}
in which the mass matrix $M_{ab}$ was introduced in Eq. \eqref{massterm}. Now we are ready to  decompose the variable $ Q^a$ into the directions along and orthogonal to time evolution~\cite{Achucarro:2012sm} as follow
\begin{equation}
\label{eq:decomposition2}
Q^a = Q^a_\bot + \dot\phi^a_0\tilde \pi \, ,
\end{equation}
with the orthogonality condition  $G_{ab} \dot{\phi}_{0}^{a}Q^b_\bot=0$.   
We impose $\tilde \pi=0$ in the comoving gauge. It is worth mentioning that, the $\tilde \pi$ mode is the fluctuation in the direction of the time translation. Furthermore, the orthogonal modes, $Q^a_\bot$, are gauge invariant quantities and are generally called ``isocurvature'' modes \cite{Gordon:2000hv}. %\textcolor{red}{
 The Mukhanov-Sasaki variable can be introduced as
\cite{gong2017multi} 
\begin{equation}
\tilde{Q}^{a}\equiv Q^{a}-\frac{\dot{\phi}_{0}^a}{H} \psi=Q^a_\bot - \frac{\dot\phi^a_0}{H}(\psi- H\tilde \pi )  \equiv Q^a_\bot - \frac{\dot\phi^a_0}{H}  \pi,
\end{equation} 
or equivalently 
\begin{equation}\label{quan}
Q^{a}\equiv  Q^a_\bot-\frac{\dot\phi^a_0}{H} ( \pi- \psi) \, .
\end{equation}
In the two-field case, with respect to the orthogonality condition, the mode $Q_\bot^a$ is proportional to the normal vector $N^a$, \textit{i.e}.,  $Q_\bot^a \propto N^a$, and  the amplitude of $Q_\bot$ is  the isocurvature field $\mathcal{F}$ \cite{gong2017multi}. On the other hand, one can replace $\pi$ simply with the curvature perturbation $\mathcal R$ in comoving gauge ($\tilde{\pi}=0$) in which  $\psi=\mathcal R$. 
Now by plugging Eq. \eqref{quan} into the quadratic action \eqref{comovingac}, the final quadratic action reads
\begin{equation}
\label{action-comoving}
 S^{2}=\int {\rm d}^4 x  \frac{a^3}{2} \Big[sgn(\pm 1)P_{,X}\Big(\dot{\mathcal{F}}^2-\frac{1}{a^2} (\partial \mathcal{F})^2\Big)-M_{NN}^2 \mathcal{F}^2- \frac{2 \mu^2\dot{\theta}}{  H}  \mathcal{F} \dot{\mathcal{R}}+\frac{M_{p}^2 \dot{H}}{  H^2 a^2} (\partial \mathcal{R})^2\Big] \, .
\end{equation}
Obviously,  the curvature perturbation $\mathcal{R}$ does not propagate in this setup. 
On the other hand,  the signature of the metric $G_{ab}$ determines the stability of perturbations. 
Therefore, the isocurvature mode does not suffer from ghost and gradient instabilities with an Lorentzian signature ($sgn(-1)=-1$), whereas in the case of the Euclidean manifold with $sgn(+1)=1$, the isocurvature perturbation is pathological.

  \section{Hamiltonian analysis of the multi-field Cuscuton gravity}\label{sec5}
This section is devoted to the full non-linear Hamiltonian analysis of the system for counting  
the correct number of degrees of freedom (DOFs).  Thus, we use the Arnowitt-Deser-Misner (ADM) decomposition \cite{arnowitt2008republication} for performing the non-linear Hamiltonian analysis. 
ADM decomposition is used to characterize the nature of gravity as a constrained system.  
The metric components of spacetime take the following form in ADM formalism
\begin{eqnarray} && \label{metricinv0}
g_{00}=-\mathcal{N}^{2}+\beta_{i}\beta^{i},\hspace{7mm} g_{0i}=\beta_{i},\hspace{7mm} g_{ij}=\gamma_{ij},\label{a8} \nonumber\\ &&
g^{00}=-\frac{1}{\mathcal{N}^{2}},\hspace{7mm} g^{0i}=\frac{\beta^{i}}{\mathcal{N}^{2}},\hspace{7mm}g^{ij}=\gamma^{ij}-\frac{\beta^{i}\beta^{j}}{\mathcal{N}^{2}},\label{a9}
\end{eqnarray}
where  $\mathcal{N}$ is the lapse function and $\beta^{i}$ is the shift vector. The spatial component of metric $\gamma_{ij}$ is defined on the three-dimensional spatial hypersurface  embedded in the full spacetime. 
In the ADM formalism, the action \eqref{22} can be taken as
\begin{equation}\label{eq:ADMaction0}
S=S_{G}+S_{M},
\end{equation}
where $S_{G}$ is related to the pure gravity part, \textit{i.e},
\begin{equation}
\label{eq:actionGR0}
S_{G} = \int d^4 x  \mathcal N\sqrt{\gamma}    \frac{M_P^2}{2}  \left( R^{(3)}+K_{ij}K^{ij}-K^2 \right)  \, ,
\end{equation}
where $R^{(3)}$ is the curvature of three-dimensional spatial hypersurface, associated with $\gamma_{ij}$.  In addition,  the extrinsic curvature, $K_{ij}$ is given by
\begin{equation}
\label{eq:extrinsiccurvature10}
K_{ij} = \frac{1}{2\mathcal N} \left( \partial_t\gamma_{ij} - \beta_{i;j} - \beta_{j;i} \right) \, , \hspace{0.5cm} K \equiv K^i{}_i \, ,
\end{equation} 
with the covariant derivative defined by the spatial metric $\gamma_{ij}$. Additionally, the matter part of the action \eqref{22} is  
\begin{equation}
S_{M}=\int  d^4 x  \mathcal N\sqrt{\gamma} \big(-\mu^{2}\sqrt{2X} -V(\phi^{a})\big),
\end{equation}
where $ X $ has the following form in ADM formalism
\begin{equation}
X=\frac{1}{2}(G_{ab}\nabla_l\phi^{a}\nabla_l\phi^{b}-\gamma^{ij}G_{ab}\partial_i \phi^{a}
\partial_j\phi^{b}),
\end{equation}
%\begin{equation}
%S=\int d^4 x \mathcal N \sqrt{\gamma} P(\frac{1}{2}(\nabla_n^2\phi^{a}-h^{ij}\partial_i \phi^{a}
%\partial_j\phi^{a})+\lambda F(\frac{1}{2}(\nabla_n^2\phi^{a}-h^{ij}\partial_i \phi^{a}
%\partial_j\phi^{a}),
%\ 
%\end{equation}
where
\begin{equation}
\nabla_l \phi^{a}=\frac{1}{\mathcal{N}}(\partial_t\phi^{a}-\beta^i\partial_i
\phi^{a}) \ .
\end{equation}
The total multi-field Cuscuton action in ADM decomposition can be expressed as
\begin{eqnarray} \label{actionH0m}&&
S=\int d^{4}x \Big[ \Pi^{ij}\partial_{t}\gamma_{ij}+\Pi^{a}\partial_{t}\phi_{a}-\mathcal{N}(\mathcal{H}_{G}+\mathcal{H}_{M})-\beta^{i}({\mathcal{H}_{G}}_{i}+{\mathcal{H}_{M}}_{i})\Big],
\end{eqnarray}
Note that  the time derivative of $\mathcal N$and $\beta^{i}$ do not exist in the action. It means that these phase space variables are not  dynamical. Therefore the dynamical variables are $\gamma_{ij}$ and $\phi^{a}$. The momentum  conjugate  of  $\gamma_{ij}$ and $\phi^{a}$,  have the following forms
\begin{eqnarray} \label{actm}&&
\Pi^{a}=\frac{\mu^2 \sqrt{\gamma}}{\sqrt{2X}}\triangledown_{l}\phi^{a},
\end{eqnarray} 
\begin{eqnarray}
\Pi^{ij}&=&\frac{\delta S_{G}}{\delta \partial_{t} \gamma_{ij}}=\frac{M_P^2}{2}\sqrt{\gamma}(K^{ij}-\gamma^{ij}K),\label{c11}
\end{eqnarray}
and
%\begin{equation}
%\mathcal{H}_{G}\equiv - \frac{M_P^2}{2}\sqrt{\gamma} \ R^{(3)}-\dfrac{2}{M_P^2 \sqrt{\gamma}}\left(\frac{1}{2}\Pi^{2}-\Pi^{ij}\Pi_{ij}\right),\hspace{0.5cm} \mathcal{H}^{i}_{G}\equiv-2\Big(\partial_{j} \Pi^{ij}+\Gamma_{jk}^{i} \Pi^{jk}\Big)
%\end{equation}
\begin{eqnarray}&&
\mathcal{H}_{M}\equiv\mu^{2}\sqrt{\gamma}\sqrt{(\frac{\Pi_{a}\Pi_{b}}{\gamma \mu^{4}}+G_{ab})g^{ij}\partial_{i}\phi^{a}\partial_{j}\phi^{b}}+\sqrt{\gamma}V(\phi^{a}) \, , {\mathcal{H}_{M}}_i=\Pi_{a}\partial_{i}\phi^{a},
\end{eqnarray}
with $\Pi\equiv \Pi^{i}_{i}$. 
Furthermore,  $ \mathcal{H}_{G} $ and  $  \mathcal{H}^{i}_{G} $  associated to the gravity part.
In this situation, we have four primary constraints $(\Pi_{\mathcal{N}},\Pi_{\beta^i})\approx 0$.
By regarding such primary constraints and the action \eqref{actionH0m}, the total Hamiltonian, from the standard definition in \cite{arnowitt2008republication, dirac1964lectures}, takes the following form.
\begin{eqnarray} &&
H_{T}=\int d^{3}x  \Big[\mathcal{N}(\mathcal{H}_{G}+\mathcal{H}_{M})+\beta^{i}({\mathcal{H}_{G}}_{i}+{\mathcal{H}_{M}}_{i})+v_{\mathcal{N}} \Pi_{\mathcal{N}}+v^{i} \Pi_{i} \Big],
\end{eqnarray}
where $v_{\mathcal{N}} $ and $v^{i}  $ are Lagrange multipliers. 
Now the  consistency of  the primary constraints $\Omega_{1}\equiv \Pi_{\mathcal N}\approx 0$ and $\Gamma_{1}^{i}\equiv \Pi^{i}\approx 0$  gives the secondary constraints as follows\footnote{Note that for our case, the Dirac brackets coincide with the Poisson brackets.
}.
\begin{eqnarray} &&
\nonumber \Omega_{2}\equiv\partial_{t} \Omega_{1}=\{\Omega_{1}(\textbf x),H_{T}(\textbf y)\}=-(\mathcal{H}_{G}+\mathcal{H}_{M})\delta^{3}(\textbf x-\textbf y) \approx 0,\\&&
\Gamma_{2}^{i}\equiv\partial_{t} \Gamma_{1}^{i}=\{\Gamma_{1}^{i}(\textbf x),H_{T}(\textbf y)\}=-({\mathcal{H}_{G}}^{i}(\textbf x)+{\mathcal{H}_{M}}^{i}(\textbf x)) \delta^{3}(\textbf x-\textbf y)\approx 0 \, .
\end{eqnarray}
We should now investigate the consistency of the secondary constraints
\begin{eqnarray} &&
\nonumber \Omega_{3} \equiv \partial_{t}\Omega_{2}=\{\Omega_{2}(\textbf x),{H}_{T}(\textbf y)\}=-\mathcal{N}\{\Omega_{2}(\textbf x),\Omega_{2}(\textbf y)\}-\beta_{i}\{\Omega_{2}(\textbf x),\Gamma_{2}^{i}(\textbf y) \}\approx 0,\\&&
\Gamma_{3}^{i} \equiv \partial_{t} \Gamma_{2}^{i}= \{\Gamma_{2}^{i}(\textbf{x}),H_{T}(\textbf y) \}=-\mathcal{N}\{\Gamma_{2}^{i}(\textbf x),\Omega_{2}(\textbf y)\}-\beta_{j}\{\Gamma_{2}^{i}(\textbf x),\Gamma_{2}^{j}(\textbf y) \}\approx 0.
\end{eqnarray}
where
\begin{eqnarray} &&
\nonumber \{ \Omega_{2}(\textbf x), \Omega_{2}(\textbf y)\}= \Gamma_{2}^{i}(\textbf y)\partial_{x^{i}}\delta^{(3)}(\textbf x-\textbf y)- \Gamma_{2}^{j}(\textbf  x)\partial_{y^{j}}\delta^{(3)}(\textbf x-\textbf y)\approx 0,\\&&
\{\Omega_{2}(\textbf x),\Gamma_{2}^{i}(\textbf y) \}=-\Omega_{2} \partial_{x_{i}}\delta^{(3)}(\textbf x-\textbf y)\approx 0,\\&&
\nonumber \{\Gamma_{2}^{i}(\textbf x),\Gamma_{2}^{j}(\textbf y)\}=\Gamma_{2}^{i}(\textbf y)\partial_{{x}_{j}}\delta^{(3)}(\textbf x-\textbf y)-\Gamma_{2}^{j}(\textbf x)\partial_{y_{i}}\delta^{(3)}(\textbf x-\textbf y)\approx 0.
\end{eqnarray}
It is obvious that  the above expressions   vanish on the constraint surface. In fact, consistency of the secondary constraints $\Omega_{2}$ and $\Gamma_{2}^{i}$ determines none of the Lagrangian multipliers and do not generate any additional constraints. Moreover,   
these eight constraints  $ \Omega_{1} $, $ \Omega_{2} $, $\Gamma_{1}^{i} $, and $ \Gamma_{2}^{i} $ are all first class constraints, which are represented as the generators of diffeomorphism.
%Now the DOFs in phase space can be read off from the master formula of the constrained system as \cite{dirac1964lectures} 
%\begin{eqnarray} &&
%\text{DOF} = N-2  \#  \text{1st Class} -\#  \text{2nd Class},\label{mf0}
%\end{eqnarray}
%where $ N $ is the total number of phase space variables.
 In summary, in this model, there are 
twenty phase space variables containing $(\mathcal N, \beta^{i},\gamma_{ij}, \Pi_{\mathcal N}, \Pi^{i},\Pi_{ij})$ and $2\mathcal{M} $ total number of the conjugate pair  $ (\phi_{a}, \Pi^{a}) $. Therefore, 
the number of DOFs \cite{dirac1964lectures} is 
\begin{eqnarray}\label{DOF0}
\text{DOF}=(20+2\mathcal{M})-16=4+2\mathcal{M},
\end{eqnarray}
which corresponds to  $(4+2\mathcal M)/2$ physical degrees of freedom in the configuration space.
Here $\mathcal{M}$ indicates the  dimension of the field space manifold or the number of scalar fields. Therefore, we have $\mathcal{M}$ extra physical degree of freedom in addition to the two gravitational degrees of freedom of general relativity. This result implies that the theory does not have the so-called Ostrogradsky ghost \cite{Woodard:2015zca}. In appendix \ref{appendixa} we investigate a Hamiltonian analysis of the multi-field cuscuton model in the homogeneous limit where $\partial_{i} \phi^{a}=0$. In single field Cuscuton case, in the the homogeneous limit ($\partial_{i} \phi=0$) the extra degree of freedom (Cuscuton scalar field)is non-dynamical leading to a theory of gravity with just two tensor degrees of freedom \cite{gomes2017hamiltonian}.

\section{Summary and conclusions}\label{sec7}
In this work, we first extended the idea of Cuscuton gravity to multi-field setup with the curved
field space manifold. Then, we found a bounce solution that has no pathologies associated the violation of NEC. More precisely, at the background level, we looked for some suitable conditions on the potential function in such a way that the bounce in the contracting or expanding phase can exist in our scenario. 
After finding such a condition, we used the cosmological perturbation theory to examine the existence of ghosts and other instabilities in our model. 

At the level of perturbations, we have used the kinematic basis in which the perturbations are decomposed into the tangential and the perpendicular to the field space trajectory. In this respect, we found that the perturbation mode tangential to background trajectory in the field space manifold, namely the adiabatic mode does not have its own dynamics at both background and perturbation levels. Nonetheless, the entropy mode perpendicular to background trajectory, originated from the extra scalar field in our model, propagates with a non-zero sound speed. 
Furthermore, we proved that whether or not the entropy perturbation is pathological directly depends on the signature of the field-space metric. To be more explicit, because the would- be-wrong-sign kinetic term of the Cuscuton is not a true kinetic term owing to the degenerate nature of the phase space, our scenario is still stable. In summary, despite violating NEC, our model provides a healthy
regular bounce. 

%Furthermore, we accomplished explicit perturbation analysis in
%the spatially flat gauge for two-field cuscuton case where we detected that the perturbation mode
%perpendicular to the background trajectory in the field space manifold, $u_N$, is excited, although
%the perturbation mode tangential to the background trajectory, $u_T$,  does not  dynamical and does not propagate. Accurately, the entropy mode arises from the extra scalar field in our model, propagates
%with the sound speed equal to unity, although the sound speed for the adiabatic mode is zero. 

%Additionally, health of the entropy perturbation depends on the signature of the field-space metric $G_{ab}$ .  Hence, the entropy mode is healthy for an Euclidean signature. However,  in the case of the Lorentzian signature,  the entropy
%perturbation suffers from the gradient instabilities and the ghost.

%Moreover,  we studied quadratic action in  the comoving gauge, in which the
%stability of perturbations depends on the signature of the metric $G_{ab}$ too. Hence, for  an Lorentzian signature (sgn(-1) = -1), there are not ghost and gradient instabilities for the isocurvature mode and the isocurvature perturbation is pathological for the Euclidean manifold with sgn(+1)= 1. 

In addition, we confirmed our findings by performing
the full non-linear Hamiltonian analysis of the multi-field Cuscuton theory and calculated the
correct number of DOFs necessary to avoid the Ostrogradsky-type ghost.

\section*{Acknowledgements}
We would like to thank Hassan Firouzjahi, Ghazal Geshnizjani, Shinji Mukohyama, and Alireza Talebian for many useful comments and discussions. We also would like to acknowledge the valuable comments and suggestions of the
referee, which have improved the quality of this paper.

\appendix

\section{Is multi-field Cuscuton gravity casual? }\label{appendixa}
As mentioned, the superluminal propagation does not necessarily lead to a breakdown of causality \cite{babichev2008k}. To make it clear, following the approach proposed in Ref. \cite{Afshordi:2006ad}, we attempt to examine whether or not this theory is causal?
 Due to the underlying Lorentz symmetry, we can always go to a frame in which locally $D\Pi^{a}\wedge D\phi^b  =0$. Consequently, there is no local dynamics in our model. 
To realize the collapse of the phase space
structure, let us change the phase space from $D\phi^{a}\wedge D\Pi^{b}$ to $D\phi^{a}\wedge D\dot{\phi}^{b}$ by the following Jacobian matrix.
 \begin{eqnarray}  &&\det\frac{\partial(\phi^{a}({\bf x})\Pi^{a}({\bf
 		x'}))}{\partial(\phi^{b}({\bf y})\dot{\phi}^{b}({\bf
 		y'}))}=\\
 &&		\det\left(\begin{array}{cc}
\delta^{a}_{b} \delta^{(3)}({\bf x -y })& 0\\ \nonumber
X^{-1}\gamma^{ij}G_{ab} \Pi^{a} \partial_{i}\phi^{b}({\bf x'}) \partial_{j}\delta^{(3)}({\bf x'-y})& \frac{-\mu^2 G_{ab}\gamma^{ij}\partial_{i}\phi^{a}({\bf x'})\partial_{j}\phi^{b}({\bf x'})}{X^{3/2}}\delta^{(3)}({\bf
 	x'-y'})\end{array}\right).\end{eqnarray}
Because the Lorentz symmetry permits us to locally rotate
 $\partial_{i}\phi^{b}({\bf x'})=0$ (the homogeneous limit) for vectors $\partial_{\mu} \phi^{a}$ such that $X=- G_{ab} \partial_{\mu} \phi^{a}\partial^{\mu} \phi^{b}/2 > 0$, the symplectic structure of the phase space collapses without carrying any local dynamical degrees of freedom. It means that local perturbations do not carry any microscopic information. Therefore, the causality can not be violated even in the presence of the superluminal sound speed.
  
%\subsection{Hamiltonian analysis in the homogeneous limit}\label{sec6}
Let us now examine the above result through performing a Hamiltonian analysis of the multi-field Cuscuton model in the homogeneous limit where $\partial_{i} \phi^{a}=0$. With respect to the action \eqref{22}, the  multi-field Cuscuton action in ADM decomposition can be expressed as
\begin{eqnarray} \label{actionHam}
S=\int d^{4}x \Big[ \Pi^{ij}\partial_{t}\gamma_{ij}+\Pi_{a}\partial_{t}\phi^{a}-\mathcal{N}(\mathcal{H}_{G}+\mathcal{H}_{M})-\beta^{i}({\mathcal{H}_{G}}_{i}+{\mathcal{H}_{M}}_{i})-\lambda^{i}_{a}\partial_i \phi^a\Big]
\end{eqnarray}
%\begin{eqnarray} \label{actm}&&
%\Pi^{a}=\mu^2 \sqrt{\gamma}X^{-1/2}\triangledown_{n}\phi^{a},
%\end{eqnarray}  
Here we have imposed the gauge fixing term $ \partial_i \phi^a  $  to the canonical Hamiltonian with a Lagrange multiplier $ \lambda^{i}_{a} $\footnote{The set up we used here is practically similar to the one applied to type-II minimally modified gravity theories in \cite{de2020theory}}.
The other Hamiltonian functions are given by
%\begin{equation}
%\mathcal{H}_{G}\equiv - \frac{M_P^2}{2}\sqrt{\gamma} \ R^{(3)}-\dfrac{2}{M_P^2 \sqrt{\gamma}}\left(\frac{1}{2}\Pi^{2}-\Pi^{ij}\Pi_{ij}\right),\hspace{0.5cm} \mathcal{H}^{i}_{G}\equiv-2\Big(\partial_{j} \Pi^{ij}+\Gamma_{jk}^{i} \Pi^{jk}\Big)
%\end{equation}
\begin{eqnarray}&&
\mathcal{H}_{M}\equiv\mu^{2}\sqrt{\gamma}\sqrt{(\frac{\Pi_{a}\Pi_{b}}{\gamma \mu^{4}}+G_{ab})g^{ij}\partial_{i}\phi^{a}\partial_{j}\phi^{b}}+\sqrt{\gamma}V(\phi^{a}) \, , {\mathcal{H}_{M}}_i=\Pi_{a}\partial_{i}\phi^{a}
\end{eqnarray}
where $\Pi\equiv \Pi^{i}_{i}$ and   $\Pi^{a}$, $ \mathcal{H}_{G} $, $  \mathcal{H}^{i}_{G} $ introduced in Section \ref{sec5}. Now, we have $ 4+3\mathcal{M} $ primary constraints  $(P_{\mathcal{N}},P_{\beta^i},\partial_i \phi^a )\approx 0$.
Thus, the total Hamiltonian function from the standard definition in \cite{bojowald2010canonical} is as follows,
\begin{equation}
\label{eq:action00}
H_{T}=\mathcal{N}(\mathcal{H}_{M}+\mathcal{H}_{G})+\beta^{i}(\mathcal{H}_{Mi}+\mathcal{H}_{Gi})+\lambda^{i}_{a}\partial_{i}\phi^{a}+u_{\mathcal{N}}P_{\mathcal{N}}+u^{i}_{\beta}{P_{i}}_{\beta} \, ,
\end{equation}
where $ u_{\mathcal{N}} $, $ \lambda^{i}_{a} $ and $ u^{i}_{\beta} $  are Lagrange multipliers which enforce the primary constraints.

To identify the secondary constraints, we should check the consistency of the primary constraints using the Poisson brackets \cite{dirac1964lectures}. 
%For gravity and matter parts, the Poisson bracket is defined, respectively, as
%\begin{eqnarray}&&\label{poisson}
%\{\mathcal{X}, \mathcal{Y}\}\equiv \int d^{3}x \Big[\frac{\delta\mathcal{X} }{\delta \gamma_{ij}(x)}\frac{\delta\mathcal{Y} }{\delta \Pi^{ij}(x)}-\frac{\delta\mathcal{Y} }{\delta \gamma_{ij}(x)}\frac{\delta\mathcal{X} }{\delta \Pi^{ij}(x)} \Big]+\Big[\frac{\delta\mathcal{X} }{\delta \phi^{a}(x)}\frac{\delta\mathcal{Y} }{\delta \Pi_{a}(x)}-\frac{\delta\mathcal{Y} }{\delta \phi^{a}(x)}\frac{\delta\mathcal{X} }{\delta \Pi_{a}(x)} \Big]\nonumber\\&&\hspace{30mm}
%\Big[\frac{\delta\mathcal{X} }{\delta \mathcal{N}(x)}\frac{\delta\mathcal{Y} }{\delta P_{\mathcal{N}}(x)}-\frac{\delta\mathcal{Y} }{\delta \mathcal{N}(x)}\frac{\delta\mathcal{X} }{\delta P_{\mathcal{N}}(x)} \Big]+\Big[\frac{\delta\mathcal{X} }{\delta \beta_{i}(x)}\frac{\delta\mathcal{Y} }{\delta P^{i}_{\beta}(x)}-\frac{\delta\mathcal{Y} }{\delta \beta_{i}(x)}\frac{\delta\mathcal{X} }{\delta P_{\beta}^{i}(x)} \Big].\nonumber
%\end{eqnarray}
Thus, the time evolution of the primary constraints are obtained to be
\begin{eqnarray}&&
\Upsilon\equiv\partial_{t}{P}_{\mathcal{N}}=-\mathcal{H}_{M}-\mathcal{H}_{G},\\&&
\Xi^{i} \equiv \partial_{t}{P}_{\beta}^{i}=-\mathcal{H}^{i}_{M}-\mathcal{H}^{i}_{G},\\&&
\Theta^{a}_{i}\equiv\partial_{t}({\partial_{i}\phi^{a}})=\mathcal{N } \mathcal{C}^{a}\partial_{i}\delta(x-y)=-(\mathcal{C}^{a}\partial_{i}\mathcal{N}+\mathcal{N}\partial_{i}\mathcal{C}^{a})\delta(x-y)\approx 0,\label{lk}
\end{eqnarray}
where
\begin{eqnarray}&&
\mathcal{C}^{a}\equiv\{\partial_{i}\phi^{a},\mu^{2}\sqrt{\gamma}\sqrt{(\frac{\Pi_{c}\Pi_{d}}{\gamma \mu^{4}}+G_{cd})g^{kl}\partial_{k}\phi^{c}\partial_{l}\phi^{d}}+V(\phi^{c})\}\nonumber\\&&=\mu^{-2}{\gamma}^{-1/2}((\frac{\Pi_{c}\Pi_{d}}{\gamma \mu^{4}}+G_{cd})g^{kl}\partial_{k}\phi^{c}\partial_{l}\phi^{d})^{-1/2}{\delta^{a}_{c}\Pi_{d}g^{kl}\partial_{k}\phi^{c}\partial_{l}\phi^{d}}.
\end{eqnarray}
Now, we should consider consistency of secondary constraints. Hence, 
$ P_{\beta}^{i} $ and $\Xi^{i}$ are first class constraints which imply that the model is invariant under spatial diffeomorphism. The consistency of the secondary constraint  $\Theta^{a}_{i}  $ gives
\begin{equation}
\varrho^{a}_{i}\equiv\partial_{t}{\Theta^{a}_{i}}=-\{\Theta^{a}_{i}, H_T\}=\varrho^{a}_{i}(u_{\mathcal{N}}, \gamma_{ij}, \Pi^{ij}, \Pi^{a}, \phi_{a})\delta(x-y)\approx 0\, ,
\end{equation}
which leads to determine the Lagrangian multiplier  $ u_{\mathcal{N}} $.  In addition, $ \mathcal{N} $  is given through phase space variables from Eq.  \eqref{lk}. Furthermore, the constraint $ \Upsilon $ satisfies the below consistency relation.
\begin{equation}
\vartheta \equiv\dot{\Upsilon}=-(\mathcal{C}^{a}\partial_{i}\lambda^{i}_{a}+\lambda^{i}_{a}\partial_{i}\mathcal{C}^{a})\delta(x-y)\approx 0\ \label{ss} ,
\end{equation}
which helps us to find $ \lambda^{i}_{a} $.  As a consequence of the above consistency relations, $ P_{\mathcal{N}} $, $ \Upsilon $, $\partial_{i}\phi^{a}  $ and $ \Theta^{a}_{i} $ are all second class constraints. 
%Hence, the dimension of  phase space reduce where degrees of freedom become 
%$ (20+2 \mathcal{M})-2\mathcal{M}=20 $. 
Thus, let us now substitute these constraints strongly, \textit{i.e.}, $P_{\mathcal{N}}=0$ and $\partial_{i}\phi^{a}=0 $ into the total Hamiltonian \eqref{eq:action00}. Additionally, the $\mathcal{N}$ function can be given by making use of Eq. \eqref{lk} and then it can be substituted into the total Hamiltonian. In this respect, the new momentum canonically  conjugate to $\phi^{a}$ is obtained form the corresponding reduced Hamiltonian as follows.  
\begin{equation} 
\tilde{\Pi}^{a}=\mu^{2}\sqrt{\gamma } (\dot{\phi}^c\dot{\phi}_c)^{-1/2}\dot{\phi}^{a}, 
\end{equation}
Note that $\tilde \Pi_{a}$ is defined to be $\tilde \Pi_{a}=G_{ab} \tilde \Pi^{b}$.
%\begin{equation} 
%\Pi^{a}\Pi^{b}=\mu^{4}\gamma \frac{(\dot{\phi}^a\dot{\phi}^b)}{\dot{\phi}^c\dot{\phi}_c}, \label{jj0}
%\end{equation}
%then multiply Eq. (\ref{jj0}) by $G_{ab}$
%\begin{equation} 
%\Pi^{a}\Pi_{a}=\mu^{4}\gamma \frac{G_{ab}(\dot{\phi}^a\dot{\phi}^b)}{\dot{\phi}^c\dot{\phi}_c}, 
%\end{equation}
Therefore, one can derive a new primary constraint $ \mathcal{A} $ as 
\begin{equation} 
\mathcal{A} \equiv \tilde \Pi^{a} \tilde \Pi_{a}-\mu^{4}\gamma \approx 0,
\end{equation}
Consequently, the reduced total Hamiltonian becomes
\begin{equation}
\label{eq:action0}
H^{R}_{T}=\mathcal{N}(\mathcal{H}_{G}+V(\phi^{a}))+\beta^{i}\mathcal{H}_{Gi}+\tau \mathcal{A} +u^{i}_{\beta}{P_{i}}_{\beta} \, ,
\end{equation}
where $ \tau $ is a new Lagrange multiplier. Note that $\mathcal{N}$ was previously given by Eq. \eqref{lk}. Allow us now to consider consistency of primary constraints $ { P_{i}}_{\beta} $ and $ \mathcal{A} $ which yield
\begin{eqnarray}&&
{\Xi^{i}}^{R} \equiv \partial_{t}{P}_{\beta}^{i}=-\mathcal{H}^{i}_{G}\approx 0,\\&&
\varTheta^{R} \equiv \partial_{t}\mathcal{A}  =\{\mathcal{A} , \mathcal{N}(\mathcal{H}_{G}+V(\phi^{a}))\}=\mathcal{D}\approx 0,
\end{eqnarray}
Clearly, $ {\Xi^{i}}^{R} $ and $ P^i_{\beta} $ are first class constraints. Moreover, $\mathcal{D} $  is a secondary constraint which involves phase space variables. As shown in the below identity, the Lagrange multiplier $ \tau $ is obtained from
the time evolution of the secondary constraint $\mathcal{D} $. 
\begin{eqnarray}&&
\partial_{t}\mathcal{D}  =\{\mathcal{D} , \mathcal{N}(\mathcal{H}_{G}+V(\phi^{a}))\}+\tau \{\mathcal{D},\mathcal{A}\}\approx 0,
\end{eqnarray}  
therefore, $ \varTheta^{R} $ and $ \mathcal{A} $ are both second class constraints. 
With the completion of these steps, we are now ready to count degrees of freedom. One can find that the total number of phase space variables is 
$ N=20+2 \mathcal{M} $ and the constraints 
$\varTheta^{R},\mathcal{A}, P_{\mathcal{N}}, \Upsilon, \partial_{i}\phi^{a}, \Theta^{a}_{i} $ are all second class constraints. About the  Lagrange multiplier  $ \lambda^{i}_{a} $, we count  only $ \mathcal{M} $ not $3 \mathcal{M}$ degrees of freedom, because using integral by parts technique,  $ \lambda^{i}_{a}\partial_{i}\phi^{a} $ can be shown to be proportional to $ \lambda_{a}\triangledown^{2}\phi^{a} $. Note that, without loss of generality, we have decomposed $ \lambda^{i}_{a} $ to scalar and vector parts, \textit{i.e.} $ \lambda^{i}_{a}=\partial^{i}{\lambda_{a}}_{S}+{\lambda^{i}_{a}}_{V}$ during this calculation. In addition, the  divergenceless  condition  $\partial_{i}\lambda^{i}_{a}=0  $ was assumed \cite{de2020theory}.

Furthermore, we take into account $\partial_{i}\phi^{a} $ and $ \Theta^{a}_{i} $ as $ 2\mathcal{M}  $ second class constraints. The constraint $ {\Xi^{i}}^{R} $ and $ P^i_{\beta} $ are also six first class constraints generating spatial diffeomorphism. 
As a result of the above discussion, 
the total number of DOFs is 
\begin{eqnarray}\label{DOF1}
\text{DOF}=(20+2\mathcal{M})-12-2-2-2\mathcal{M}=4,
\end{eqnarray}
which implies the system has two physical degrees of freedom in the configuration space. Namely, this finding confirms that the multi-field Cuscuton theory in the homogeneous limit adds no additional degrees of freedom to a gravitational system.

\vspace{1cm}
%=====================================================================
\bibliographystyle{JHEP}
\bibliography{BibMimetic1}

\providecommand{\href}[2]{#2}\begingroup\raggedright\begin{thebibliography}{10}

\bibitem{Horndeski:1974wa}
G.~W. Horndeski, \emph{{Second-order scalar-tensor field equations in a
  four-dimensional space}},
  \href{https://doi.org/10.1007/BF01807638}{\emph{Int. J. Theor. Phys.}
  {\bfseries 10} (1974) 363--384}.

\bibitem{Deffayet:2011gz}
C.~Deffayet, X.~Gao, D.~A. Steer and G.~Zahariade, \emph{{From k-essence to
  generalised Galileons}},
  \href{https://doi.org/10.1103/PhysRevD.84.064039}{\emph{Phys. Rev. D}
  {\bfseries 84} (2011) 064039},
  [\href{https://arxiv.org/abs/1103.3260}{{\ttfamily 1103.3260}}].

\bibitem{Zumalacarregui:2012us}
M.~Zumalacarregui, T.~S. Koivisto and D.~F. Mota, \emph{{DBI Galileons in the
  Einstein Frame: Local Gravity and Cosmology}},
  \href{https://doi.org/10.1103/PhysRevD.87.083010}{\emph{Phys. Rev. D}
  {\bfseries 87} (2013) 083010},
  [\href{https://arxiv.org/abs/1210.8016}{{\ttfamily 1210.8016}}].

\bibitem{Gao:2011qe}
X.~Gao and D.~A. Steer, \emph{{Inflation and primordial non-Gaussianities of
  'generalized Galileons'}},
  \href{https://doi.org/10.1088/1475-7516/2011/12/019}{\emph{JCAP} {\bfseries
  12} (2011) 019}, [\href{https://arxiv.org/abs/1107.2642}{{\ttfamily
  1107.2642}}].

\bibitem{Gleyzes:2014dya}
J.~Gleyzes, D.~Langlois, F.~Piazza and F.~Vernizzi, \emph{{Healthy theories
  beyond Horndeski}},
  \href{https://doi.org/10.1103/PhysRevLett.114.211101}{\emph{Phys. Rev. Lett.}
  {\bfseries 114} (2015) 211101},
  [\href{https://arxiv.org/abs/1404.6495}{{\ttfamily 1404.6495}}].

\bibitem{Gleyzes:2014qga}
J.~Gleyzes, D.~Langlois, F.~Piazza and F.~Vernizzi, \emph{{Exploring
  gravitational theories beyond Horndeski}},
  \href{https://doi.org/10.1088/1475-7516/2015/02/018}{\emph{JCAP} {\bfseries
  02} (2015) 018}, [\href{https://arxiv.org/abs/1408.1952}{{\ttfamily
  1408.1952}}].

\bibitem{Nicolis:2008in}
A.~Nicolis, R.~Rattazzi and E.~Trincherini, \emph{{The Galileon as a local
  modification of gravity}},
  \href{https://doi.org/10.1103/PhysRevD.79.064036}{\emph{Phys. Rev. D}
  {\bfseries 79} (2009) 064036},
  [\href{https://arxiv.org/abs/0811.2197}{{\ttfamily 0811.2197}}].

\bibitem{Afshordi:2006ad}
N.~Afshordi, D.~J.~H. Chung and G.~Geshnizjani, \emph{{Cuscuton: A Causal Field
  Theory with an Infinite Speed of Sound}},
  \href{https://doi.org/10.1103/PhysRevD.75.083513}{\emph{Phys. Rev. D}
  {\bfseries 75} (2007) 083513},
  [\href{https://arxiv.org/abs/hep-th/0609150}{{\ttfamily hep-th/0609150}}].

\bibitem{afshordi2007cuscuton}
N.~Afshordi, D.~J. Chung, M.~Doran and G.~Geshnizjani, \emph{Cuscuton
  cosmology: dark energy meets modified gravity},
  \href{https://doi.org/10.1103/PhysRevD.75.123509}{\emph{Physical Review D}
  {\bfseries 75} (2007) 123509}.

\bibitem{Afshordi:2009tt}
N.~Afshordi, \emph{{Cuscuton and low energy limit of Horava-Lifshitz gravity}},
  \href{https://doi.org/10.1103/PhysRevD.80.081502}{\emph{Phys. Rev. D}
  {\bfseries 80} (2009) 081502},
  [\href{https://arxiv.org/abs/0907.5201}{{\ttfamily 0907.5201}}].

\bibitem{bhattacharyya2018revisiting}
J.~Bhattacharyya, A.~Coates, M.~Colombo, A.~E.
  G{\"u}mr{\"u}k{\c{c}}{\"u}o{\u{g}}lu and T.~P. Sotiriou, \emph{Revisiting the
  cuscuton as a lorentz-violating gravity theory},
  \href{https://doi.org/10.1103/PhysRevD.97.064020}{\emph{Physical Review D}
  {\bfseries 97} (2018) 064020}.

\bibitem{hovrava2009quantum}
P.~Ho{\v{r}}ava, \emph{Quantum gravity at a lifshitz point},
  \href{https://doi.org/10.1103/PhysRevD.79.084008}{\emph{Physical Review D}
  {\bfseries 79} (2009) 084008}.

\bibitem{ito2019dressed}
A.~Ito, A.~Iyonaga, S.~Kim and J.~Soda, \emph{Dressed power-law inflation with
  a cuscuton}, \href{https://doi.org/10.1103/PhysRevD.99.083502}{\emph{Physical
  Review D} {\bfseries 99} (2019) 083502}.

\bibitem{de2017caustics}
C.~de~Rham and H.~Motohashi, \emph{Caustics for spherical waves},
  \href{https://doi.org/10.1103/PhysRevD.95.064008}{\emph{Physical Review D}
  {\bfseries 95} (2017) 064008}.

\bibitem{adshead2016magnon}
P.~Adshead, D.~Blas, C.~Burgess, P.~Hayman and S.~P. Patil, \emph{Magnon
  inflation: slow roll with steep potentials},
  \href{https://doi.org/10.1088/1475-7516/2016/11/009}{\emph{Journal of
  Cosmology and Astroparticle Physics} {\bfseries 2016} (2016) 009}.

\bibitem{bartolo2022cuscuton}
N.~Bartolo, A.~Ganz and S.~Matarrese, \emph{Cuscuton inflation},
  \href{https://doi.org/10.1088/1475-7516/2022/05/008}{\emph{Journal of
  Cosmology and Astroparticle Physics} {\bfseries 2022} (2022) 008}.

\bibitem{DeFelice:2022uxv}
A.~De~Felice, K.-i. Maeda, S.~Mukohyama and M.~C. Pookkillath, \emph{{VCDM and
  Cuscuton}},  \href{https://arxiv.org/abs/2204.08294}{{\ttfamily 2204.08294}}.

\bibitem{de2020theory}
A.~De~Felice, A.~Doll and S.~Mukohyama, \emph{A theory of type-ii minimally
  modified gravity},
  \href{https://doi.org/10.1088/1475-7516/2020/09/034}{\emph{Journal of
  Cosmology and Astroparticle Physics} {\bfseries 2020} (2020) 034}.

\bibitem{boruah2018cuscuton}
S.~S. Boruah, H.~J. Kim, M.~Rouben and G.~Geshnizjani, \emph{Cuscuton bounce},
  \href{https://doi.org/10.1088/1475-7516/2018/08/031}{\emph{Journal of
  Cosmology and Astroparticle Physics} {\bfseries 2018} (2018) 031}.

\bibitem{quintin2020cuscuton}
J.~Quintin and D.~Yoshida, \emph{Cuscuton gravity as a classically stable
  limiting curvature theory},
  \href{https://doi.org/10.1088/1475-7516/2020/02/016}{\emph{Journal of
  Cosmology and Astroparticle Physics} {\bfseries 2020} (2020) 016}.

\bibitem{kim2021spectrum}
J.~L. Kim and G.~Geshnizjani, \emph{Spectrum of cuscuton bounce},
  \href{https://doi.org/10.1088/1475-7516/2021/03/104}{\emph{Journal of
  Cosmology and Astroparticle Physics} {\bfseries 2021} (2021) 104}.

\bibitem{dubovsky2006null}
S.~Dubovsky, T.~Gr{\'e}goire, A.~Nicolis and R.~Rattazzi, \emph{Null energy
  condition and superluminal propagation},
  \href{https://doi.org/10.1088/1126-6708/2006/03/025}{\emph{Journal of High
  Energy Physics} {\bfseries 2006} (2006) 025}.

\bibitem{rubakov2014null}
V.~A. Rubakov, \emph{The null energy condition and its violation},
  \href{https://doi.org/10.3367/UFNe.0184.201402b.0137}{\emph{Physics-Uspekhi}
  {\bfseries 57} (2014) 128}.

\bibitem{sawicki2013hidden}
I.~Sawicki and A.~Vikman, \emph{Hidden negative energies in strongly
  accelerated universes},
  \href{https://doi.org/10.1103/PhysRevD.87.067301}{\emph{Physical Review D}
  {\bfseries 87} (2013) 067301}.

\bibitem{libanov2016generalized}
M.~Libanov, S.~Mironov and V.~Rubakov, \emph{Generalized galileons:
  instabilities of bouncing and genesis cosmologies and modified genesis},
  \href{https://doi.org/10.1088/1475-7516/2016/08/037}{\emph{Journal of
  Cosmology and Astroparticle Physics} {\bfseries 2016} (2016) 037}.

\bibitem{kobayashi2016generic}
T.~Kobayashi, \emph{Generic instabilities of nonsingular cosmologies in
  horndeski theory: A no-go theorem},
  \href{https://doi.org/10.1103/PhysRevD.94.043511}{\emph{Physical Review D}
  {\bfseries 94} (2016) 043511}.

\bibitem{Ijjas:2016vtq}
A.~Ijjas and P.~J. Steinhardt, \emph{{Fully stable cosmological solutions with
  a non-singular classical bounce}},
  \href{https://doi.org/10.1016/j.physletb.2016.11.047}{\emph{Phys. Lett. B}
  {\bfseries 764} (2017) 289--294},
  [\href{https://arxiv.org/abs/1609.01253}{{\ttfamily 1609.01253}}].

\bibitem{Ijjas:2016tpn}
A.~Ijjas and P.~J. Steinhardt, \emph{{Classically stable nonsingular
  cosmological bounces}},
  \href{https://doi.org/10.1103/PhysRevLett.117.121304}{\emph{Phys. Rev. Lett.}
  {\bfseries 117} (2016) 121304},
  [\href{https://arxiv.org/abs/1606.08880}{{\ttfamily 1606.08880}}].

\bibitem{Creminelli:2012my}
P.~Creminelli, K.~Hinterbichler, J.~Khoury, A.~Nicolis and E.~Trincherini,
  \emph{{Subluminal Galilean Genesis}},
  \href{https://doi.org/10.1007/JHEP02(2013)006}{\emph{JHEP} {\bfseries 02}
  (2013) 006}, [\href{https://arxiv.org/abs/1209.3768}{{\ttfamily 1209.3768}}].

\bibitem{Dobre:2017pnt}
D.~A. Dobre, A.~V. Frolov, J.~T. G\'alvez~Ghersi, S.~Ramazanov and A.~Vikman,
  \emph{{Unbraiding the Bounce: Superluminality around the Corner}},
  \href{https://doi.org/10.1088/1475-7516/2018/03/020}{\emph{JCAP} {\bfseries
  03} (2018) 020}, [\href{https://arxiv.org/abs/1712.10272}{{\ttfamily
  1712.10272}}].

\bibitem{Easson:2013bda}
D.~A. Easson, I.~Sawicki and A.~Vikman, \emph{{When Matter Matters}},
  \href{https://doi.org/10.1088/1475-7516/2013/07/014}{\emph{JCAP} {\bfseries
  07} (2013) 014}, [\href{https://arxiv.org/abs/1304.3903}{{\ttfamily
  1304.3903}}].

\bibitem{Easson:2011zy}
D.~A. Easson, I.~Sawicki and A.~Vikman, \emph{{G-Bounce}},
  \href{https://doi.org/10.1088/1475-7516/2011/11/021}{\emph{JCAP} {\bfseries
  11} (2011) 021}, [\href{https://arxiv.org/abs/1109.1047}{{\ttfamily
  1109.1047}}].

\bibitem{boruah2017theory}
S.~S. Boruah, H.~J. Kim and G.~Geshnizjani, \emph{Theory of cosmological
  perturbations with cuscuton},
  \href{https://doi.org/10.1088/1475-7516/2017/07/022}{\emph{Journal of
  Cosmology and Astroparticle Physics} {\bfseries 2017} (2017) 022}.

\bibitem{iyonaga2018extended}
A.~Iyonaga, K.~Takahashi and T.~Kobayashi, \emph{Extended cuscuton:
  formulation},
  \href{https://doi.org/10.1088/1475-7516/2018/12/002}{\emph{Journal of
  Cosmology and Astroparticle Physics} {\bfseries 2018} (2018) 002}.

\bibitem{gong2011covariant}
J.-O. Gong and T.~Tanaka, \emph{A covariant approach to general field space
  metric in multi-field inflation},
  \href{https://doi.org/10.1088/1475-7516/2011/03/015}{\emph{Journal of
  Cosmology and Astroparticle Physics} {\bfseries 2011} (2011) 015}.

\bibitem{gong2017multi}
J.-O. Gong, \emph{Multi-field inflation and cosmological perturbations},
  \href{https://doi.org/10.1142/S021827181740003X}{\emph{International Journal
  of Modern Physics D} {\bfseries 26} (2017) 1740003}.

\bibitem{gong2016path}
J.-O. Gong, M.-S. Seo and G.~Shiu, \emph{Path integral for multi-field
  inflation}, \href{https://doi.org/10.1007/JHEP07(2016)099}{\emph{Journal of
  High Energy Physics} {\bfseries 2016} (2016) 1--31}.

\bibitem{cespedes2012importance}
S.~Cespedes, V.~Atal and G.~A. Palma, \emph{On the importance of heavy fields
  during inflation},
  \href{https://doi.org/10.1088/1475-7516/2012/05/008}{\emph{Journal of
  Cosmology and Astroparticle Physics} {\bfseries 2012} (2012) 008}.

\bibitem{achucarro2017cumulative}
A.~Ach{\'u}carro, V.~Atal, C.~Germani and G.~A. Palma, \emph{Cumulative effects
  in inflation with ultra-light entropy modes},
  \href{https://doi.org/10.1088/1475-7516/2017/02/013}{\emph{Journal of
  Cosmology and Astroparticle Physics} {\bfseries 2017} (2017) 013}.

\bibitem{gomes2017hamiltonian}
H.~Gomes and D.~C. Guariento, \emph{Hamiltonian analysis of the cuscuton},
  \href{https://doi.org/10.1103/PhysRevD.95.104049}{\emph{Physical Review D}
  {\bfseries 95} (2017) 104049}.

\bibitem{langlois2008perturbations}
D.~Langlois and S.~Renaux-Petel, \emph{Perturbations in generalized multi-field
  inflation},
  \href{https://doi.org/10.1088/1475-7516/2008/04/017}{\emph{Journal of
  Cosmology and Astroparticle Physics} {\bfseries 2008} (2008) 017}.

\bibitem{Elliston:2012ab}
J.~Elliston, D.~Seery and R.~Tavakol, \emph{{The inflationary bispectrum with
  curved field-space}},
  \href{https://doi.org/10.1088/1475-7516/2012/11/060}{\emph{JCAP} {\bfseries
  11} (2012) 060}, [\href{https://arxiv.org/abs/1208.6011}{{\ttfamily
  1208.6011}}].

\bibitem{babichev2008k}
E.~Babichev, V.~Mukhanov and A.~Vikman, \emph{k-essence, superluminal
  propagation, causality and emergent geometry},
  \href{https://doi.org/10.1088/1126-6708/2008/02/101}{\emph{Journal of High
  Energy Physics} {\bfseries 2008} (2008) 101}.

\bibitem{trofimov2002riemannian}
V.~V. Trofimov and A.~T. Fomenko, \emph{Riemannian geometry}, {\emph{Journal of
  Mathematical Sciences} {\bfseries 109} (2002) 1345--1501}.

\bibitem{petrov1969phys}
A.~Petrov, \emph{Phys. einstein spaces},  1969.

\bibitem{muller1997closed}
U.~M{\"u}ller, C.~Schubert and A.~van~de Ven, \emph{A closed formula for the
  riemann normal coordinate expansion}, {\emph{arXiv preprint gr-qc/9712092}
  (1997) }.

\bibitem{Gordon:2000hv}
C.~Gordon, D.~Wands, B.~A. Bassett and R.~Maartens, \emph{{Adiabatic and
  entropy perturbations from inflation}},
  \href{https://doi.org/10.1103/PhysRevD.63.023506}{\emph{Phys. Rev. D}
  {\bfseries 63} (2000) 023506},
  [\href{https://arxiv.org/abs/astro-ph/0009131}{{\ttfamily
  astro-ph/0009131}}].

\bibitem{Mansoori:2021fjd}
S.~A.~H. Mansoori, A.~Talebian, Z.~Molaee and H.~Firouzjahi, \emph{{Multifield
  mimetic gravity}},
  \href{https://doi.org/10.1103/PhysRevD.105.023529}{\emph{Phys. Rev. D}
  {\bfseries 105} (2022) 023529},
  [\href{https://arxiv.org/abs/2108.11666}{{\ttfamily 2108.11666}}].

\bibitem{Achucarro:2012sm}
A.~Achucarro, J.-O. Gong, S.~Hardeman, G.~A. Palma and S.~P. Patil,
  \emph{{Effective theories of single field inflation when heavy fields
  matter}}, \href{https://doi.org/10.1007/JHEP05(2012)066}{\emph{JHEP}
  {\bfseries 05} (2012) 066},
  [\href{https://arxiv.org/abs/1201.6342}{{\ttfamily 1201.6342}}].

\bibitem{arnowitt2008republication}
R.~Arnowitt, S.~Deser and C.~W. Misner, \emph{Republication of: The dynamics of
  general relativity},
  \href{https://doi.org/10.1007/s10714-008-0661-1}{\emph{General Relativity and
  Gravitation} {\bfseries 40} (2008) 1997--2027}.

\bibitem{dirac1964lectures}
P.~Dirac, \emph{Lectures on quantum mechanics (new york: Belfer graduate school
  of science, yeshiva university press)}, .

\bibitem{Woodard:2015zca}
R.~P. Woodard, \emph{{Ostrogradsky's theorem on Hamiltonian instability}},
  \href{https://doi.org/10.4249/scholarpedia.32243}{\emph{Scholarpedia}
  {\bfseries 10} (2015) 32243},
  [\href{https://arxiv.org/abs/1506.02210}{{\ttfamily 1506.02210}}].

\bibitem{bojowald2010canonical}
M.~Bojowald, \emph{Canonical gravity and applications: cosmology, black holes,
  and quantum gravity}.
\newblock Cambridge University Press, 2010.

\end{thebibliography}\endgroup
%\bibliographystyle{JHEP}
%=====================================================================

\end{document}